\documentclass[reqno, 12pt]{article}
\usepackage{a4wide,amssymb}
\usepackage{amsthm,amsfonts,amsmath}
\usepackage{graphicx,xcolor
,
}
\usepackage{eucal}
\usepackage{enumerate}
\usepackage[normalem]{ulem}
\usepackage{mathtools}

\newcommand{\R}{{\mathbb R}}         
 
\newcommand{\E}{{\mathbb E}}

\renewcommand{\d}{\delta}
           
\newcommand\om{\omega}

\def\G{\Gamma}

\def\L{{\Lambda}}

\def\TTT{{\mathbb{T}}}

\def\MM{{\cal M}}

\newtheorem{thm}{Theorem}[section]

\newtheorem{lem}[thm]{Lemma}

\newtheorem{cor}[thm]{Corollary}

\def\be{\begin{equation}}
\def\ee{\end{equation}}
\def\bea{\begin{eqnarray}}
\def\eea{\end{eqnarray}}
\def\ni{\noindent}
\def\nn{\nonumber}

\def\d{\delta}
\def\o{\omega}
\def\b{\beta}
\def\t{\tau}

\usepackage{multicol}
\usepackage{xpatch}
\makeatletter
\xapptocmd\thenomenclature{\let\@item\nomencl@item\def\nomencl@width{0pt}}{}{}
\let\nomencl@item\@item
\xpretocmd\nomencl@item{\nomencl@measure{#1}}{}{}
\def\nomencl@measure#1{%
  \sbox0{#1}%
  \ifdim\wd0>\nomencl@width\relax
    \E^Ndef\nomencl@width{\the\wd0}%
  \fi
}

\xpatchcmd\thenomenclature
  {\section*{\nomname}}
  {\begin{multicols}{2}[\section*{\nomname}]}
  {}{}
\xpatchcmd\thenomenclature
  {\chapter*{\nomname}}
  {\begin{multicols}{2}[\chapter*{\nomname}]}
  {}{}

\xapptocmd\endthenomenclature{%
  \immediate\write\@mainaux{\global\nomlabelwidth\nomencl@width\relax}%
  \end{multicols}
}{}{}
\makeatother

\begin{document}

\begin{titlepage}

\begin{center} 
{\bf \Large{On the cardinality of collisional clusters for hard spheres at low density
}}

\vspace{1.5cm}
{\large M. Pulvirenti$^{1}$ and S. Simonella$^{2}$}

\vspace{0.5cm}
{$1.$ \scshape {\small Dipartimento di Matematica, Universit\`a di Roma La Sapienza\\ 
Piazzale Aldo Moro 5, 00185 Rome -- Italy,  and \\
 International Research Center M\&MOCS, Universit\`a dell'Aquila, \\ Palazzo Caetani, 04012 Cisterna di Latina -- Italy.}  
 \smallskip

$2.$ {\small UMPA UMR 5669 CNRS, ENS de Lyon \\ 46 all\'{e}e d'Italie,
69364 Lyon Cedex 07 -- France}}

\end{center}

\vspace{1.5cm}

\vspace{1.5cm}
\noindent
{\bf Abstract.} 
We resume the investigation, started in \cite{Aoki}, of the statistics of backward clusters in a gas of $N$ hard spheres of small diameter $\varepsilon$. A backward cluster is defined as the group of particles involved directly or indirectly in the backwards-in-time dynamics of a given tagged sphere. We obtain an estimate of the average cardinality of clusters with respect to the equilibrium measure, global in time, uniform in $\varepsilon, N$ for $\varepsilon^2 N=1$ (Boltzmann-Grad regime).

\vspace{0.5cm}\noindent
{\bf Keywords.} Hard spheres; low--density gas; Boltzmann equation; backward cluster.

\thispagestyle{empty}

\bigskip
\bigskip

\newpage

\end{titlepage}


\section{Introduction} \label{sec:intro}
\setcounter{equation}{0}    
\def\theequation{1.\arabic{equation}}
Boltzmann's equation, its actual relation with particle systems and higher order density corrections have led, through the years, to 
several mathematical questions around collisions in hard-sphere type models. A famous example is 
the problem of counting the number of collisions in a group of hard balls \cite{MC00}. This is of interest in its own and it is a problem of great geometric complexity \cite{Va79,BFK98,Se20}. More closely related to the kinetic theory of gases is the analysis of the statistical (average) behaviour of collision sequences, for a finite subgroup of particles in a large (eventually infinite) system of hard spheres. At low density, if particle ``$1$'' undergoes $k$ collisions, it will typically interact with $k$ different spheres. Collecting these ``fresh'' particles, together with the fresh particles encountered by them, and so on, we arrive to a natural notion of ``cluster of influence'', to be associated to the particle $1$. These clusters have both theoretical and applied interest and they are involved in the control of dynamical correlations \cite{Aoki,PulvirSimonBG,BGS-RS20}. 
Our goal is to derive an estimate on the cardinality of collisional clusters, uniform in the low-density (Boltzmann-Grad) limit. The estimate is valid globally in time, both with respect to an equilibrium measure, or with respect to a nonequilibrium measure with strong boundedness properties. Note that the control of the size of backward clusters implies the absence of very large chains of collisions, which could prevent the propagation of chaos in the Boltzmann-Grad limit.

The definition of cluster is given in the next section. Our main result (Theorem \ref{thm}) is presented in Section \ref{secMR}, while Section \ref{subsec:heur} discusses the difficulties encountered in controlling the size of clusters.  Section \ref{sec:proof} is devoted to the proof of the theorem.

In this paper we consider hard-sphere systems, however we believe that our results can be extended to particle systems interacting via suitable (say, repulsive) short-range potentials, by using approach and techniques inspired from \cite{Gr58,Ki75,GSRT12E, PSS}.

\subsection{Backward and forward clusters}

Consider a system of $N$ 
identical hard spheres of diameter $\varepsilon>0$ moving in the three-dimensional space. 
To fix the ideas we will confine the spheres to the torus $\TTT^3=[0,2\pi)^3$ (although different boundary conditions might be considered).
The  spheres collide by means of  the laws of elastic reflection \cite{Alexander}.
A configuration of the system is $Z_N=( z_1, \cdots, z_N )$, where $z_i=(x_i,v_i) \in \TTT^{3} \times \R^3$ are the position 
and the velocity of particle $i$  respectively. 

Given a particle, say particle $1$, consider $z_1 (t, Z_N)$ its state 
(position and velocity) at time $t$  for the initial configuration $Z_N$.
We define the {\em backward cluster of particle $1$} (at time $t$ and for the initial configuration $Z_N$)
as the ordered set of particles with indices  $BC(1) \subset I_N:=\{ 1,2, \cdots , N\}$,  constructed in the following way.
Going back in time starting from $z_1 (t, Z_N)$, let $i_1$ be the  first particle colliding with $1$. Next, considering the two particles $1$ and $i_1$,  let us go back in time up to the first collision of one particle of the pair with a new particle $i_2$. We iterate this procedure up to time $0$.
Then  $BC(1) :=\{ i_1,i_2, \cdots , i_n \}$ with $i_r \neq i_s$ for $r \neq s$.

In \cite{Aoki} we introduced and studied the problem of determining the time evolution of $|BC(1)|$, the cardinality of backward clusters. We showed that this hard task simplifies considerably in the Boltzmann-Grad limit \cite{Gr49}, as one can resort to the Boltzmann equation to prove exponential growth.
We also discussed the connection with the hierarchy of equations. At low density, dynamical observables admit a perturbative representation as `sums over backward clusters'. This is suggested, in particular, by a special representation of the solution to the Boltzmann equation, well known under the name of Wild sum \cite{Wild, Aoki}.

We point out here that a definition of {\em forward cluster} $FC(1)$ can be given by the same identical procedure, just reversing the direction in time. Namely we start from $z_1$ (configuration of particle $1$ at time zero) and move up to $z_1 (t, Z_N)$, drawing the forward cluster of influence. This notion may have a different applied interest as e.g.\,in the rate of spread of an epidemic \cite{PSv20}. In the present paper, we will focus on the backward cluster, although our results will be obviously true for forward clusters as well.

\begin{figure}[htbp] 
\centering
\includegraphics[width=5in]{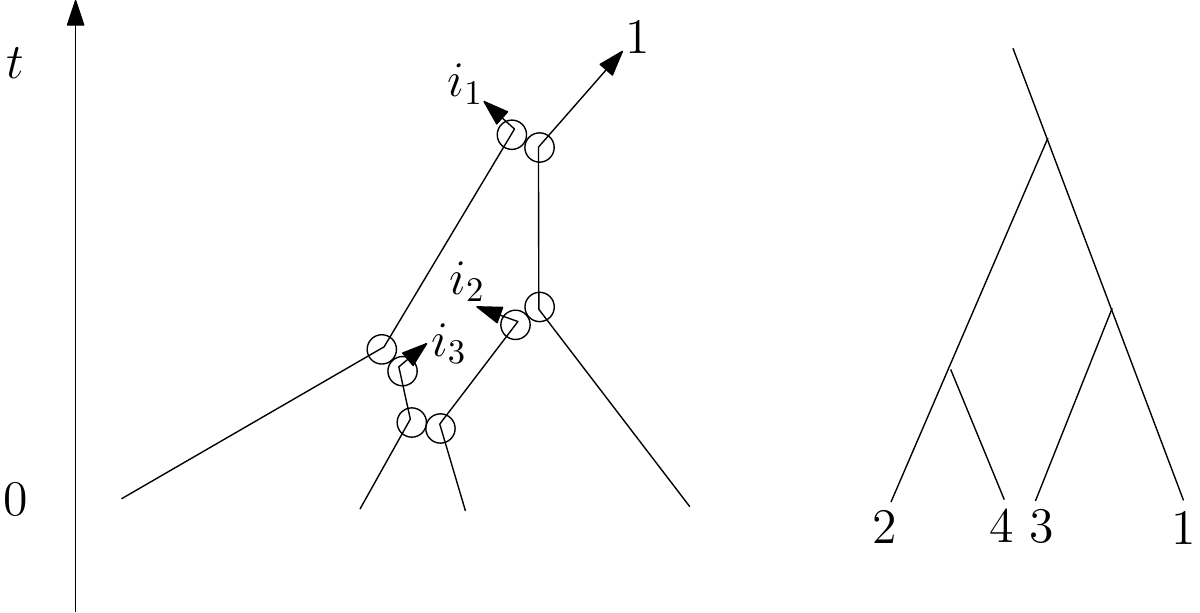} 
\caption{The trajectory of a backward cluster $BC(1)$ at time $t$ (of cardinality $3$) is represented on the left. Its tree structure $\G_3 = (1,1,2)$ is given by the graph on the right.}
\label{fig:treedef}
\end{figure}

We may add to the previous definition an internal {\em structure}, by specifying the sequence of colliding pairs. To do this, we introduce binary tree graphs. A {\em $n-$collision tree} $\G_n$ is the 
 collection of integers $k_1,\cdots,k_n$ such that
\be \label{eq:1pTR}
k_1\in I_1, k_2 \in I_{2}, \cdots, k_n\in I_{n}\;,\ \ \ \ \ \ \mbox{where\ \ \ \ \ \ $I_s=\{1,2,\cdots,s\}$\;.}
\ee
We say that the backward cluster  $BC(1)$ has structure $\G_n=(k_1, \cdots, k_{n})$ if $|BC(1)| = n$ and
 the ordering of the collisions producing the cluster is specified by the tree.
 For a graphical representation, see the example in Figure \ref{fig:treedef}.

Clearly, the collisions defining the backward cluster (involving a ``new'' particle) are not the only collisions showing up
in the trajectory of the cluster. Collisions which do not involve a new particle are called {\em recollisions} (e.g.\;the collision between $(i_2,i_3)$ in the figure).

Let now assign a  probability density $W_0^N$ symmetric in the exchange of particles, on the $N$--particle phase space, 
\be \label{eq:PhSp}
\MM_N := \Big\{ Z_N \in  \left(\TTT^{3} \times \R^3\right)^N,\ \ |x_i-x_j| > \varepsilon,\ \ i\neq j\Big\}\; .
\ee
In general, the measure is non-stationary and we denote by $W^N(t)$ the time-evolved density. This is transported along the hard-sphere flow $Z_N \to \Phi_N^t(Z_N)$, $t>0$, precisely defined as follows. Given a time--zero configuration $Z_N \in \MM_N$, each particle will move on a straight line with constant velocity between collisions;
when two hard spheres collide with positions $x_i, x_j$ at distance $\varepsilon$, normalized relative distance 
$\o = (x_i-x_j)/|x_i-x_j|=(x_i-x_j)/\varepsilon \in {\mathbb S}^2$ and incoming velocities $v_i, v_j$ (i.e. $(v_i-v_j)\cdot\o <0$), 
these are instantaneously transformed to outgoing velocities $v'_i, v'_j$ (i.e. $(v'_i-v'_j)\cdot\o >0$) through 
the relations 
\be
\begin{cases}
\displaystyle v_i'=v_i-\om [\om\cdot(v_i-v_j)] \\
\displaystyle  v_j'=v_j+\om[\om\cdot(v_i-v_j)]
\end{cases}\;.\label{eq:coll}
\ee
We recall that $\Phi_N^t$ is a.e.\;well defined with respect to the Lebesgue measure \cite {Alexander, Va79, CIP}.

 For $j=1,2,\cdots,N$, we denote by $f^{N}_{0,j}$ and $f^{N}_{j} (t) $ the 
$j-$particle marginals of $ W_0^N$ and $W^N(t)$ respectively (e.g.\,$f^{N}_{j} (t) 
= \int W^N(t) \,dz_{j+1}\cdots dz_N$).
The main quantity of interest is then
\be
\label{epsgamma}
f^{N, \G_n}_1 (z_1, t) = \int dz_2 \cdots dz_N \, \chi _{\G_n}\,W^N(Z_N,t)\,
\ee
where $\chi _{\G_n}$ is the characteristic function of the event: {\it Particle $1$ has a backward cluster of cardinality $n$ with structure $\G_n $}. Eq.\;\eqref{epsgamma} is the restriction of the marginal $f^N_1$ to trajectories with given cluster structure,
hence $f^N_1= \sum_n f^{N,n}_1 $ where
 \be \label{eq:fndef}
f^{N,n}_1 =   \sum_{\G_n}f^{N, \G_n}_1\;.
 \ee
The average cluster size at time $t$ is
\be
\label{S}
S^N(t)= \E^N (|BC(1)|) =\sum_{k = 0}^{N-1}\, k\, \mathbb{P}^N \left(|BC(1)|=k\right)
\ee
where
\be
\label{P}
\mathbb{P}^N \left(|BC(1)|=k\right)=\sum_{\G_k}\int dz_1\, f^{N, \G_k}_1(z_1, t)\;.  
\ee
We shall look for estimates of these quantities, uniformly in the Boltzmann-Grad scaling
$$N \to \infty\;,\quad \varepsilon \to  0$$ 
where
\be \label{eq:BG}
\varepsilon^2 N=1\;.
\ee
Hence $N$ is the only relevant parameter: from now on, $\varepsilon$ is determined by the relation \eqref{eq:BG}.

We conclude this section by mentioning still a different definition of cluster which has been studied considerably.
Fixed a time $t>0$, one can partition the entire system of particles into the maximal disjoint union of non-interacting (in $[0,t]$) groups. Such groups are `dynamical clusters' (called in \cite{PSW1} {\em Bogolyubov clusters}), generally bigger than  the  backward clusters of  the  particles composing them. Indeed in the backward cluster, the trajectory of each particle is specified only in a subinterval of $(0, t)$, see e.g.\;Figure \ref{fig:treedef}; while the dynamical cluster ``completes''  the future history of particles $i_1, i_2, \cdots$, together with the complete history (in $[0,t]$) of the particles with whom they collide, and so on: 
$$
\includegraphics[width=3.3in]{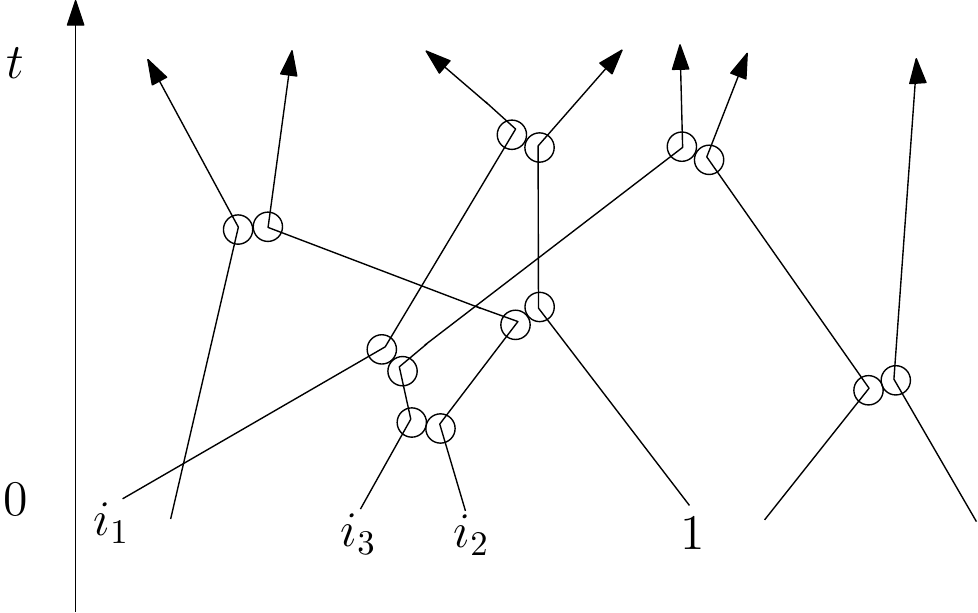} 
$$
\noindent This leads to a {\em symmetrized} notion of cluster which does not depend on the direction of time \cite{PSW1}. Theoretical investigation of dynamical clusters has been performed in \cite{Sinai1,Sinai2,GKBSZ08,PSW1,PSW2}. They have an interesting 
geometric property as they give rise to ``percolation" in finite time (close to the mean free time in the Boltzmann-Grad limit). That is, a giant, macroscopic cluster (of size $O(N)$) emerges abruptly, so that after some critical time it is impossible to obtain an estimate of the mean size uniformly in $N$.

\section{Main results}  \label{secMR}
\setcounter{equation}{0}    
\def\theequation{2.\arabic{equation}}

Let the initial probability measure $W_0^N$ on the $N$-particle hard sphere system be given by 
 the canonical Gibbs measure at temperature $\b^{-1}>0$:
 \be
\label{eq}
W^{N}_{eq} (Z_N) := \frac 1 {{ \cal Z}_N} \prod_{i=1}^N e^{-\frac{\b}{2} v_i^2}\;,
\ee
where ${ \cal Z}_N$ is the normalization constant. Remind that this is a probability measure on the phase space \eqref{eq:PhSp}, which takes into account the exclusion condition $|x_i -x_j | >\varepsilon$, $i \neq j$, with $\varepsilon$ given by \eqref{eq:BG}.
Moreover, the measure is time invariant for the hard-sphere flow.

Let $f^{N, \G_n}_{1,eq}$, $f^{N, n}_{1,eq}$, $S^N_{eq}$, $\mathbb{P}^N_{eq}$ be given as in \eqref{epsgamma}-\eqref{P} for the equilibrium measure $W^N_{eq}$.

\begin{thm} \label{thm}
Given $t>0$, there exist an integer $k_0$ and a positive constant $C$ such that
the cardinality of backward clusters at time $t$ satisfies, for $k>k_0$, the following inequality:
\be
\label{res1}
\mathbb{P}^N_{eq} (|BC(1)|=k)\leq  C t\,e^{-\frac 14 k^{\frac 1 {Ct}}}\;.
\ee
In particular, the average cardinality $S^N_{eq} (t)$ is bounded uniformly in $N$ in any bounded interval of time. 
\end{thm}

The integer $k_0$ deteriorates for $t/\sqrt\b$ large, as expected. Moreover, the bound implies
superexponential growth $S^N_{eq} (t) = O(e^{C't\log t})$, $C'>0$ for $t$ large, with $C' \sim 1/\sqrt\b$ for $\b$ small. This is compatible with the results of \cite{Aoki}.

The same estimate holds true out of equilibrium, if one assumes strong uniform bounds on the marginals.
\begin{cor} \label{cor}
Let $W^N(t)$ have marginals $\left(f^{N}_{j}(t)\right)_{j=1}^N$ obeying
\be
\label{bound}
f^{N}_{j}(t) \prod_{i=1}^j e^{\frac{\b}{2} v_i^2} \leq A^{ j}\;,\quad t \in [0,T]
\ee
for some $A,\b, T >0$.
Then the cardinality of backward clusters at time $t \in [0,T]$ satisfies Eq.\,\eqref{res1}.
\end{cor}
The proof of the corollary is exactly the same as the proof of the theorem, because only the bound \eqref{bound} is used as property of the equilibrium marginals.
As a consequence Eq.\,\eqref{res1} is proven under the assumptions of \cite{Lanford}, for times short enough (in fact as proved by Lanford,  \eqref{bound} holds true uniformly in the Boltzmann-Grad scaling).

Our results, combined with the results on the Boltzmann equation obtained in \cite{Aoki}, suggest that the average growth of clusters is indeed exponential in time, at least for measures on which we have a very good control.

\subsection{A heuristic bound} \label{subsec:heur}

The probability that particle $1$ has a backward cluster of cardinality $k$ at time $t$ can be estimated as follows. Suppose that all the particles have velocity of order $1$. Some particle $i_1$ has to lie in the collision cylinder spanned by 1 in $(0,t)$, which corresponds to a region of volume
$C (N-1) \,\varepsilon^2\, t $ in the $N$-particle phase space, where $C>0$ is a geometrical constant. Similarly, the condition of existence of a second particle $i_2$ restricts to a volume of order $ 2 C (N-1)(N-2)  \,\varepsilon^4\, t^2 $ (because of the two possibilities: $i_2$ can be ``generated'' by particle $1$ or by particle $i_1$).
Iterating and using  \eqref{eq:BG} we find that 
\be\label{eq:Rb}
\mathbb{P}^N (|BC(1)|=k) \leq \frac {\left(C\,t\right)^k\, k! } {k!} \;,
\ee
where the $k!$ at denominator arises from the time ordering which we have to  take into account.

Eq.\,\eqref{eq:Rb} is a rough bound, plausible for short times only. In fact the argument reminds Lanford's proof of the local validity of the Boltzmann equation \cite{Lanford} (see also \cite{IP89,Spohn,CIP,GSRT12E,PSS,PulvirSimonBG,De17,Bodi,GH21}). 
The estimate is too pessimistic, because it ignores that the particles not belonging to $BC(1)$ do {\em not} interact with it, which produces exponential damping. 

A simple way to provide an improved, formal estimate is the following.
Fix $\G_k$ and focus on $\int f^{N, \G_k}_1(t)$, 
the probability that particle $1$ has a backward cluster of cardinality $
k$ with structure $\G_k$.
We partition the interval $(0,t)$ into $M$ disjoint small time intervals of length $\d = t/M$. Then
\be
\label{expr}
\int f^{N, \G_k}_1\approx \sum_{\substack { m_1, \cdots, m_k \\ m_s=1,\cdots,M \\ M \geq  m_1>m_2, \cdots, m_k\geq 1} } \,
\int f^{N, \G_k}_1\,\chi_{m_1,\cdots,m_k}
\ee
where $\chi_{m_1,\cdots,m_k}$ is the indicator of the event: {\em the $k$ collisions of the backward cluster (recollisions excluded) take place in the time intervals $((m_s-1) \d,  m_s \d)$, $s=1,\cdots,k$}.
Eq.\;\eqref{expr} is not exact because we are assuming that in each time interval only one new particle appears, an error which we neglect in view of the limit $\d \to 0$. As before, the probability of the collisions in the small time intervals is approximately $N\varepsilon^2 \d=\d $, but now we take into account the probability of the complement set $1-\d$ (no collision generating new particles takes place in the other intervals). Thus the first collision of particle $1$ in $((m_1-1) \d,  m_1 \d)$ yields a contribution
$
 \d (1- \d)^{M-m_1},
$
(taking into account that in the time interval $(m_1\d, t)$ particle $1$ did not collide). The second collision of the tree yields
$
\d^2 (1-\d)^{M-m_1} (1-\d)^{2(m_1-m_2)} (1-\d)^{-2}
$,
and so on. In conclusion
\bea
\int f^{N, \G_k}_1\approx && \sum_{\substack { m_1, \cdots, m_k \\ m_s=1,\cdots,M \\ M \geq  m_1>m_2, \cdots, m_k\geq 1} }
\frac{ \d^k}{(1-\d)^{\frac {k(k+3)}2}}
(1-\d )^{(M-m_1)} (1- \d )^{2(m_1-m_2)} \cdots (1- \d )^{(k+1) m_k} \nn \\
\approx && \int_0^t dt_1 \cdots \int_0^{t_{k-1}} dt_k \,\, e^{- (t-t_1)} e^{- 2(t_1-t_2)}\cdots 
e^{-k(t_{k-1}-t_k)}e^{-(k+1)t_k }= e^{- t}  \frac {(1-e^{- t})^k}{k!} \,\,  \nn
\eea
as $\d \to 0$. 
Indeed if $t_i=m_i\d$:
$$
(1-\d)^{(i+1) (m_i-m_{i+1})} =\left(1-\frac tM \right)^{( i+1) (t_i-t_{i+1} ) \frac Mt} \approx e^{ -(i+1) (t_i-t_{i+1} ) }.
$$
The reader might recognize the Riemann approximation of the term with structure $\G_k$ in the classical
Wild sum (see e.g.\;\cite{Aoki}, Section 3). 
Summing over all  the trees (see \eqref{P}), we get
$$
 \mathbb{P}^N(|BC(1)|=k) \approx e^{- t} (1-e^{- t})^k\;,
 $$
and the average size $
 S^N(t) \approx e^{t}-1$ is bounded for all positive times.
  
In essence, the above computation is equivalent to replacing the (difficult) computation of $  f^{N, \G_n} (t)$, 
with the computation of the `limiting quantity' as given by a nice solution to the Boltzmann equation.
Such a quantity is provided by the Wild sums which can be estimated in the case of homogeneous solutions. This has been done in fact in \cite{Aoki}, and compared with molecular dynamics simulations, to argue 
that $S^N(t)$ should increase exponentially in time:
$$
\left(e^{ct}-1\right) \leq S^N(t) \leq \left(e^{Ct}-1\right)
$$
for $t$ large, for some $c, C>0$.

\section{Proof of Theorem \ref{thm}} \label{sec:proof}
\setcounter{equation}{0}    
\def\theequation{3.\arabic{equation}}

The proof is organized in two parts. In Section \ref{subsec:prel} we introduce a formula expressing
integrals over paths of a given cluster, and give a first a priori bound on \eqref{eq:fndef}.
This is inspired by previous works on the Boltzmann-Grad limit. In Section \ref{subsec:est}
we prove \eqref{res1}, by controlling the dynamical representation on small, properly chosen time intervals obtained by partitioning $(0,t)$.

Throughout the proof, $\chi_E$ will indicate the characteristic function of the set $E$.

\subsection{Integrals over clusters} \label{subsec:prel}

We start by introducing a notation for the trajectory of a backward cluster. 
For future convenience, we shall look at clusters on a fixed time interval $[t^*,t]$ where $0 \leq t^*<t$.
We still denote by $BC(1)$ the backward cluster of 1, only in this subsection restricted to $[t^*,t]$.
According to a terminology introduced in \cite{PulvirSimonBG}, the trajectory of such a cluster is called {\em interacting backwards flow} (IBF). We denote it by $Z^{IBF} (s) $, $s \in [t^*,t]$.
Note that there is no label specifying the number of particles. This number depends indeed on the time, as explained by the following construction. 

Given $\G_k$, we fix a collection of variables $z_1 $, $T_k$, $\Omega_k$, $V_{1,k}$:
\bea
&& z_1 \in \TTT^{3} \times \R^3\;,\nn\\
&& T_k = \left(t_1,\cdots,t_k\right)\in \R^k\;,\nn\\
&& \Omega_k = \left(\o_1,\cdots,\o_k\right)\in  {\mathbb S}^{2k}\;,\nn\\
&& V_{1,k} = \left(v_{2},\cdots,v_{1+k}\right)\in \R^{3k}\nn
\eea
where the times are constrained to be ordered as
$$t\equiv t_0 > t_1 > t_2 > \cdots > t_k > t_{k+1}\equiv t^*\;,$$
and $\Omega_k$ has to satisfy a further constraint that will be specified soon.
If $s \in (t_{r+1},t_r)$, then the IBF contains exactly $1+r$ 
particles:
\be
Z^{IBF} (s) = (z^{IBF}_1(s),\cdots,z^{IBF}_{1+r}(s)) \in \MM_{1+r}\ \ \ \ \ \mbox{for }s \in (t_{r+1},t_r)\;, \nn
\ee
with
\be
z^{IBF}_i (s)=(\xi_i(s),\eta_i(s))\;,\nn
\ee
respectively the position and the velocity of particle $i$ in the flow.
The trajectory $Z^{IBF} (s)$ is constructed starting from the configuration $z_1$ at time $t>0$, and going back in time.
A new particle appears (is ``created'') at time $t_r$ in a collision state with a previous particle $k_r \in\{1, \cdots, r \}$, specified by $\G_k$.
More precisely, in the time interval $(t_r,t_{r-1})$ particles $1,\cdots,r$ flow according to the interacting dynamics
of $r$ hard spheres.
 This defines $Z^{IBF}_{r}(s)$ starting from $Z^{IBF}_{r}(t_{r-1}).$ At time $t_r$ the 
particle $1+r$ is created by particle $k_r$ in the position $\xi_{1+r}(t_r)= \xi_{k_r}(t_r)+ \omega_r\,\varepsilon$
and with velocity $v_{1+r}$. This defines $Z^{IBF}(t_r) = (z^{IBF}_1(t_r),\cdots,z^{IBF}_{1+r}(t_r)).$
A constraint on $\o_r$ is imposed ensuring that two hard spheres cannot be at distance smaller than $\varepsilon$.
Next, the evolution
in $(t_{r+1},t_r)$ is contructed applying to this configuration the dynamics of $1+r$ spheres.
Since, by construction, $\o_r \cdot (v_{1+r}-\eta_{k_r}(t_r^+)) \geq 0$, the 
pair is post--collisional. Then the presence of the interaction in the hard-sphere flow forces the pair to perform a 
(backwards) instantaneous collision. Proceeding inductively, the IBF is constructed for all times $s\in [t^*,t]$.

Observe that, to denote the explicit dependence on the whole set of variables, we should write
\be
Z^{IBF}(s) = Z^{IBF}\left(s; \G_k,  z_1, T_k, \Omega_k, V_{1,k}\right)\;,\quad s \in [t^*,t]\;.
\label{eq:cmps}
\ee
We shall however use always the abbreviated notation.

Let ${\cal A}\left(\G_k\right) \subset \MM_N$ be the set of variables $Z_N$ such that
the trajectory $\Phi_{N}^{-(t-s)}\left(Z_{N}\right)$, $s \in [t^*,t]$, satisfies the following constraints:

\noindent (i) the backward cluster of $1$ in $[t^*,t]$ is given by $BC(1) = \{2,\cdots,1+k\}$;

\ni (ii) the structure of  $BC(1)$ is given by $\G_k$\;

\ni (iii) the backward cluster has $n'$ creations in $[t',t]$ and $k-n'$ creations in  $[t^*,t')$, for some intermediate time $t' \in [t^*,t]$.

We introduce a map ${\cal I}$ on ${\cal A}\left(\G_k\right)$ by
\be
\label {map}
{\cal I}\left(Z_{N}\right)  =  \left(z_1\,,  T_k\,, \Omega_k\,,  V_{1,k}, Z_{1+k,N-k-1}(t^*)\right)
\ee
where $(z_1\,,  T_k\,, \Omega_k\,,  V_{1,k})$ is the set of variables such that \eqref{eq:cmps} provides the trajectory of $\{1\} \cup BC(1)$, and $Z_{1+k,N-k-1}(t^*) = z_{k+2}(t^*),\cdots,z_{N}(t^*)$ is the configuration of the remaining particles at time $t^*$. The map ${\cal I}$ is invertible on ${\cal A}\left(\G_k\right)$ and its Jacobian determinant has absolute value $$\varepsilon^{2k} \prod_{i=1}^k \o_i\cdot \left(v_{1+i}-\eta_{k_i} (t_i^+)\right)$$ (positive by construction). 
To construct ${\cal I}^{-1}$, one determines first the IBF \eqref{eq:cmps} using the variables $z_1\,,  T_k\,, \Omega_k\,,  V_{1,k}$. Secondly one adds the variables $Z_{1+k,N-k-1}$, and consider $\left(Z^{IBF}(t^*),Z_{1+k,N-k-1}\right)$ as initial condition for the flow $\Phi_N^{s}$, $s \in [0,t-t^*]$. By construction, the variables external to the backward cluster do not interfere with it. Therefore by uniqueness of the hard sphere dynamics, $\Phi_N^{t}\left(Z^{IBF}(t^*),Z_{1+k,N-k-1}\right)$ is determined.

Let
$$
B_i= \left(\o_i\cdot \left(v_{1+i}-\eta_{k_i} (t_i^+)\right)\right)_+\chi_{\{|\xi_{1+i}(t_i)-\xi_{k}(t_i)| > \varepsilon\  \forall k\neq k_i\}}\;.
$$
Notice that the constraint ensures non-overlap of hard spheres in the IBF (at the creation times).
Using the Liouville equation and \eqref{map}, we have that
\bea
\label{if}
&&\int_{{\cal A}\left(\G_k\right)}dz_2\cdots dz_{N}\, W^N(Z_{N} ,t)  \nn\\
&&  = \varepsilon^{2k} \int d\Lambda_{\G_k}
\, \prod_{i=1}^k B_i \, W^N \left( Z^{IBF}_{1+k}(t^*), Z_{1+k,N-k-1},  t^*\right) 
\eea
where 
$$d\Lambda_{\G_k} = dT_k\, d\Omega_k \,dV_{1,k}\,dZ_{1+k,N-k-1}\,\chi_{\{t > t_1>t_2\cdots > t_k > t^*\}}\,
\chi_{{\cal I}\left({\cal A}\left(\G_k\right)\right)}
$$
Notice that ${\cal I}\left({\cal A}\left(\G_k\right)\right)$ is a rather intricate subset of the full domain of integration. It depends on all the variables and in particular it imposes restrictions on particles $1+i$ concerning their behaviour in $(t_i,t)$.

Ignoring the restriction to ${\cal I}\left({\cal A}\left(\G_k\right)\right)$ and using the definition of marginal, we obtain a first estimate:
\bea
\label{if'}
\int_{{\cal A}\left(\G_k\right)}dz_2\cdots dz_{N}\, W^N(Z_{N} ,t)  \leq \varepsilon^{2k} \int d\Lambda_{k}\,
\chi_{\{t_{n'} > t' > t_{n'+1}\}}
\, \prod_{i=1}^k B_i \, f^N_{1+k}(Z^{IBF}_{1+k}(t^*),t^*)\;,
\eea
where 
$$d\Lambda_{k} = dT_k\, d\Omega_k \,dV_{1,k}\,\chi_{\{t > t_1>t_2\cdots > t_k > t^*\}}\;$$
and $\chi_{\{t_{n'} > t' > t_{n'+1}\}}$ comes from condition (iii) above.

Moreover, the following uniform bound is well known. It is essentially equivalent to Lanford's a priori bound providing the short time validity result \cite{Lanford}.

 \begin{lem} \label{prop:apb}
Suppose that there exist $A,\b>0$ such that 
\be
f^N_{j}(Z_j,t^*) e^{+ \frac{\b}{2} \sum_{i=1}^jv_i^2} \leq A^j \label{eq:hypb}
\ee
for all $j =  1,\cdots,N$ and $Z_j \in \MM_j$. Then there exists $C_1= C_1(A,\b)>0$ such that, for all $k\geq 1$
and $t' \in [t^*,t]$,
\be
\label{prop}
\sum_{\G_k}\int d\Lambda_{k}\,\chi_{\{t_{n'} > t' > t_{n'+1}\}}
\, \prod_{i=1}^k B_i \, f^N_{1+k}(Z^{IBF}_{1+k}(t^*),t^*) \leq eA  C_1^k
(t-t')^{n'} (t'-t^*)^{k-n'}e^{-\frac {\b} 4 v_1^2} \;.
\ee
\end{lem}
The proof of the lemma is recalled in Appendix.

\subsection{Estimate of $|BC(1)|$} \label{subsec:est}

Given an arbitrary but fixed $t>0$, we write $t=m\t$ where $\t>0$, sufficiently small, will be chosen later on.
We partition the interval $[0,t]$ into $m$ intervals $[0,\tau)$, $[\tau,2\tau)$, $\cdots$, $[(m-1)\tau,m\tau=t]$.
The intervals are indexed by $i =1,2,\cdots,m$ respectively.
For any integer $k>0$, we assign to the $i-$th interval the number $$\kappa_i=\frac {k^{\frac {m-i+1}m}}2\;,$$ 
to be used as a cutoff on the local growth of the cluster.
The sequence is decreasing
 $$
 \kappa_1=\frac k 2\,,\  \kappa_2=\frac {k^{\frac {m-1}m}} 2\,,\,  \cdots\,,\, \kappa_m=\frac {k^{\frac 1m}}2
 $$
 and
$$
\frac {\kappa_i}k \to 0\;\;(i \neq 1)\,,  \quad   \frac {\kappa_{i+1}}{\kappa_i}=k^{-\frac 1 m} \to 0
$$ 
as $k$ diverges.

By \eqref{epsgamma}-\eqref{eq:fndef},
\bea
\label {exp1} 
f_{1,eq}^{N,k}(z_1,t)&&=  \sum_{\G_k}\int dz_2 \cdots dz_N \, \chi _{\G_k}\, W^N_{eq} \left(Z_N\right) \nn \\
&& =(N-1) \cdots (N-k) \sum_{\G_k} \int dz_2 \cdots dz_N\, \chi ^{ord}_{\G_k}\, W^N_{eq} \left(Z_N\right)\nn \\
&& = (N-1) \cdots (N-k) \sum_{ \substack { n_1 \cdots n_m : \\ \sum_{i=1}^m n_i =k }}
 \sum_{\G_k} \int dz_2 \cdots dz_N \,\chi ^{ord}_{\G_k} \,\chi_{\{n_1, \cdots, n_m\}}\,W^N_{eq} \left(Z_N\right) \;.\nn\\
\eea
Recall that $\chi _{\G_k} $ is the characteristic function of particle $1$ having a backward cluster of cardinality $k$ with structure $\G_k $. Similarly $\chi ^{ord}_{\G_k}$ is the characteristic function of particle $1$ having the backward cluster $BC(1) = \{2,\cdots,k+1\}$, with structure $\G_k $. The second identity in \eqref{exp1} is due to the symmetry of the measure.
Finally by definition, the characteristic function of the set $\{n_1, \cdots, n_m\}$ constrains the number of creations in the backward cluster in  the interval $\left[(i-1)\tau,i \tau\right)$ to be exactly $n_i$.

We insert now the partition of unity
$$
1= \sum_{s=1}^{m} \prod_{i=s+1}^m \chi_{\{n_i \leq \kappa_i\}}  \chi_{\{n_s >\kappa_s\}} +\prod_{i=1}^m \chi_{\{n_i \leq \kappa_i\}}\;.
$$
Note that $\prod_{i=1}^m \chi_{\{n_i \leq \kappa_i\}}=0$ for $k$ large enough
since
$$
\sum_{i=1}^m n_i \leq \sum_{i=1}^m \kappa_i \leq \frac { k +(m-1) k^{\frac {m-1}m}} 2 \leq \frac 23 k
$$
for $k$ large (depending on $m$), so that  condition $\sum n_i=k$ implies that there exists at least an interval for which $n_s > \kappa_s$.
Therefore 
\bea
\label {exp2} 
&& f_{1,eq}^{N,k}(z_1,t)  = (N-1) \cdots (N-k) \sum_{s=1}^m \nn \\
&&\ \ \ \times \sum_{ \substack { n_1 \cdots n_m : \\ \sum_{i=1}^m n_i =k }} \sum_{\G_k} \int dz_2 \cdots dz_N  \,W^N_{eq} \left(Z_N\right)\chi_{\{n_1, \cdots, n_m\}}  \left(\prod_{i=s+1}^m \chi_{\{n_i \leq \kappa_i\}} \right) \chi_{\{n_s >\kappa_s\}} 
\,\chi^{ord}_{\G_k}\;. \nn
\eea

We set $\bar n_s=\sum_{i=s+1}^m n_i$, $ \bar \kappa_s=\sum_{i=s+1}^m \kappa_i$ and $\chi^s_{\{\bar n_s,n_s\}}$ the indicator function of having $\bar n_s$ creations in the time interval  $[s\tau,t]$ and $n_s$ creations in the interval $[(s-1)\tau, s\tau)$.   Note then that the last line in the previous formula is bounded above by
\bea
&&  \sum_{ \bar n_s \leq \bar \kappa_s}  \sum_{ n_s > \kappa_s}\,\sum_{\G_k}\int  \,W^N_{eq} \,\chi^s_{\{\bar n_s, n_s\}}\,\chi^{ord}_{\G_k}\nn\\
&&\leq \frac{1}{(N-1-\bar n_s-n_s) \cdots (N-k)}  \sum_{ \bar n_s \leq \bar \kappa_s}  \sum_{ n_s > \kappa_s}\,\,\sum_{\G_{\bar n_s+n_s}} \int \,W^N_{eq}  \,\chi^s_{\{\bar n_s, n_s\}}\,
\chi^{ord, s}_{\G_{\bar n_s+n_s}}\;,\nn
\eea
where in the last step we eliminated the sum over all trees in $[0, (s-1)\tau)$, and $\chi^{ord, s}_{\G_{\bar n_s+n_s}}$ is the characteristic function of particle $1$ having the backward cluster $\{2,\cdots,\bar n_s+ n_s+1\}$ with structure $\G_{\bar n_s+n_s}$ in 
$[(s-1)\tau,t]$. In this way  we have removed any constraint on the trajectory in the time interval $[0, (s-1)\tau)$, thus
 \bea
\label {exp3} 
f_{1,eq}^{N,k}(z_1,t)  && \leq \sum_{s=1}^m   \sum_{ \bar n_s \leq \bar \kappa_s}  \sum_{ n_s > \kappa_s}
(N-1) \cdots (N-\bar n_s-n_s) \nn \\
&& \ \ \ \times\sum_{\G_{\bar n_s+n_s}} \int dz_2 \cdots dz_N \,W^N_{eq} \left(Z_N\right)\,\chi^s_{\{\bar n_s, n_s\}}\,\chi^{ord,s}_{\G_{\bar n_s+n_s}}   \;.\nn
\eea

Applying now \eqref{if'} (with $k\to\bar n_s+n_s$, $t^* = (s-1)\tau$, $t' = s\tau$ and $n' = \bar n_s$), we arrive to
\bea
f_{1,eq}^{N,k}(z_1,t)  
&& \leq \sum_{s=1}^m \sum_{r_1 \leq \bar \kappa_s} \sum_{r_2 >\kappa_s}\sum_{\G_{r_1+r_2}} \,\varepsilon^{2(r_1+r_2)}\,  (N-1) \cdots (N-r_1 -r_2 ) \nn \\
 &&\ \ \ \times
\int d\L_{r_1+r_2}\,\chi_{\{t_{r_1} > s\tau > t_{r_1+1}\}}\, \prod_{i=1}^{r_1+r_2} B_i\,
  f^{N}_{1+r_1+r_2,eq}\left( Z^{IBF}_{1+r_1+r_2}\left((s-1)\tau\right)\right)   \;.\nn
\eea
Since $
\varepsilon^{2(r_1+r_2)}  (N-1) \cdots (N-r_1 -r_2 ) \leq 1\;,
$
by Lemma \ref{prop:apb} we conclude that
$$
f_{1,eq}^{N,k}(z_1,t)   \leq \sum_{s=1}^m \sum_{r_1 \leq \bar \kappa_s} \sum_{r_2 >\kappa_s}
eA \,C_1^{r_1+r_2}\, t^{r_1} \,\tau^{r_2}\,  e^{-\frac {\b} 4 v_1^2}\;.
$$

We therefore choose $m = \left\lceil{2C_1t}\right\rceil$,
 hence $\tau \leq \frac{1}{2C_1}$ which implies
$$
f_{1,eq}^{N,k}(z_1,t)   \leq \sum_{s=1}^m  \,\left(C_2 t\right)^{\bar \kappa_s} \,\left(\frac{1}{2}\right)^{\kappa_s}\,  e^{-\frac {\b} 4 v_1^2}
$$
for some large enough $C_2>0$. Since
$
\bar \kappa_s\leq m\,\kappa_{s+1} =m\,k^{-\frac 1m} \kappa_s\;,
$
it follows that
$$
f_{1,eq}^{N,k}(z_1,t)   \leq \sum_{s=1}^m \, \left[ \frac{1}{2}\left(C_2 t\right)^{m\, k^{-\frac{1}{m}}}\right]^{\kappa_s}
\,  e^{-\frac {\b} 4 v_1^2}\;.
$$
The term in square brackets can be made smaller than $1/\sqrt e$ for $k$ large enough, hence 
\eqref{res1} is obtained after integrating in $z_1$. \qed

\addcontentsline{toc}{subsection}{Appendix. \ \ \ Proof of Lemma \ref{prop:apb}}
 \subsection*{Appendix: Proof of Lemma \ref{prop:apb}} \label{sec:E}
 \setcounter{equation}{0}    
\def\theequation{A.\arabic{equation}}

The conservation of energy at collisions implies 
\be
\sum_{i=1}^{1+k}(\eta_i(0))^2 = \sum_{i=1}^{1+k}v_i^2\;.\nn
\ee
In particular $\sum_{k_i=1}^{i}\left(\eta_{k_i}(t_i^+)\right)^2\leq\sum_{i=1}^{1+k}v_i^2$.
It follows that 
\be
\sum_{\G_k}\prod_{i=1}^k B_i \leq \prod_{i=1}^k \Big[(1+k)|v_{1+i}|+
(1+k)^{\frac{1}{2}}\Big(\sum_{l=1}^{1+k} v_l^2\Big)^{\frac{1}{2}}\Big]\;.\nn
\ee
Moreover
\bea
&& \Big(\sum_{l=1}^{1+k} v_l^2\Big)^{\frac{1}{2}} e^{-\frac {\beta }{4k}\sum_{i=1}^{1+k} v_i^2  } \leq 
\sqrt{\frac{2k}{e\b}}\;.\nn
\eea
Using the assumption \eqref{eq:hypb} in the l.h.s.\,of \eqref{prop}, the estimates above imply that we can bound it by%
\bea
 e^{-(\b/4)v_1^2} A \int dT_k\, d\Omega_k \,dV_{1,k}\,\chi_{\{t_{n'} > t' > t_{n'+1}\}} \,\prod_{i=1}^k e^{-\frac{\b}{4}v_{1+i}^2} \left((1+k)|v_{1+i}|
+\frac{\sqrt{2k(1+k)}}{\sqrt{e\b}} \right)\;.\nn\\ \label{eq:proofSTE}
\eea
The integral on the velocities factorizes so that
\be
\mbox{\eqref{eq:proofSTE}} \leq e^{-(\b/4)v_1^2} A
 C_\b^{k}\frac{(t-t')^{n'} (t'-t^*)^{k-n'}}{n'! (k-n')!} (1+k)^k\nn  \label{eq:estBproof1}
\ee
for a suitable constant $C_\b>0$
(explicitly computable in terms of gaussian integrals). Since 
\be
\frac{(1+k)^k}{n'! (k-n')!} \leq 2^k\frac{(1+k)^k}{k!} \leq 2^k\frac{(1+k)^{1+k}}{(1+k)!} \leq 2^ke^{1+k}\;,\nn \label{eq:nnnfact}
\ee
we obtain \eqref{prop}.

\bigskip\bigskip\bigskip

\noindent
{\bf Acknowledgments.} We thank the hospitality of the HIM during the Bonn Junior
Trimester Program {\em Kinetic Theory} in 2019, where this work was started.

\end{document}